\documentclass[aps,prl,twocolumn,superscriptaddress,showpacs]{revtex4}
\usepackage{graphicx}
\usepackage{epstopdf}

\begin{document}

\title{A robust platform cooled by superconducting electronic refrigerators}

\author{H. Q. Nguyen}
\affiliation{Low Temperature Laboratory (OVLL), Aalto University School of Science, P.O. Box 13500, 00076 Aalto, Finland}
\affiliation{Nano and Energy Center, Hanoi University of Science, VNU, Hanoi, Vietnam}
\author{M. Meschke}
\affiliation{Low Temperature Laboratory (OVLL), Aalto University School of Science, P.O. Box 13500, 00076 Aalto, Finland}
\author{J. P. Pekola}
\affiliation{Low Temperature Laboratory (OVLL), Aalto University School of Science, P.O. Box 13500, 00076 Aalto, Finland}

\begin{abstract}
A biased tunnel junction between a superconductor and a normal metal can cool the latter electrode. Based on a recently developed cooler with high power and superior performance, we have integrated it with a dielectric silicon nitride membrane, and cooled phonons from 305 mK down to 200 mK. Without perforation and covered under a thin alumina layer, the membrane is rigorously transformed into a cooling platform that is robust and versatile for multiple practical purposes. We discuss our results and possibilities to further improve the device.
\end{abstract}
\maketitle
Refrigeration below 1 kelvin belongs traditionally to the domain of macroscopic machines, such as adiabatic demagnetization or $^3$He-$^4$He dilution cryostats. Although being reliable and universal, these systems are often bulky and complicated to operate. Alternatively, one can employ the cooling effect in a Normal metal - Insulator - Superconductor (NIS) junction when such a contact is biased near the superconducting gap \cite{NahumAPL, MuhonenRPP, GiazottoRMP06, PekolaPRL04, LeivoAPL96, VasenkoPRB10,KoppinenPRL09,ONeilPRB12}. Powered by NIS coolers, a non-conducting platform can cool a transition edge sensor down to its working temperature \cite{MillerAPL08,VercruyssenAPL11} or it can ultimately replace a dilution unit \cite{LowellAPL13}. This approach offers a simple, light weight solution to cool mesoscopic devices such as qubits \cite{MartinisNature10}, nanomechanical resonators \cite{SillanpaaNature13}, electron pumps \cite{PekolaRMP13}, or low temperature detectors \cite{Enss}, especially those for spaceborne applications \cite{Tauber}.

There are many challenges that hinder an uncompromised demonstration of such a device. First, the SINIS cooler under-performs its theoretical prediction, mainly due to the excessive number of injected quasiparticles in the superconducting lead \cite{MillerAPL08, VercruyssenAPL11, LowellAPL13, ClarkAPL04}. On the other hand, those devices that would perform well, lack a practical power to handle external loads \cite{PekolaPRL04, KoppinenPRL09, CourtPRB08}. Second, when the device dimensions become comparable to the phonon wavelength, a few $\mu$m at low temperatures, size effects modify thermal transport properties \cite{Majumdar}, especially at the boundary of materials \cite{PohlRMP89}. It becomes increasingly difficult to precisely engineer the thermal transport, especially in the sub kelvin temperature regime. As a result, one often forces to suppress the thermal conductivity between the cold platform and its environment as effectively as possible. The platform of choice is an amorphous silicon nitride (SiN) membrane, where three dimensional phonons do not exist \cite{LeivoAPL98,HolmesAPL98}. This two dimensional structure is then typically perforated so that the center part is suspended only by thin legs \cite{LeivoAPL98,Luukanen,MillerAPL08}. It results in a fragile structure and further manipulation is next to impossible, although there are outstanding exceptions \cite{LowellAPL13}. 

Recently, we have developed a SINIS electron cooler with a well thermalized superconductor using photolithography and wet etching \cite{NguyenAPL,NguyenNJP}. This device has enough power to handle an external load, yet it performs outstandingly over a wide range of temperature \cite{NguyenLowT}. In this paper, we demonstrate an integration of these coolers on a SiN membrane. The platform reaches 200 mK from cryostat temperature of 305 mK without perforation of the membrane. With 1 nW cooling power at 300 mK, one pair of junctions is capable of cooling the membrane. We arrange three other auxiliary coolers that power other coldfingers arranged in an "onion" like configuration to shunt part of the heat from the external bath.

\begin{figure}[t]
\begin{center}
\includegraphics[width=3in,keepaspectratio]{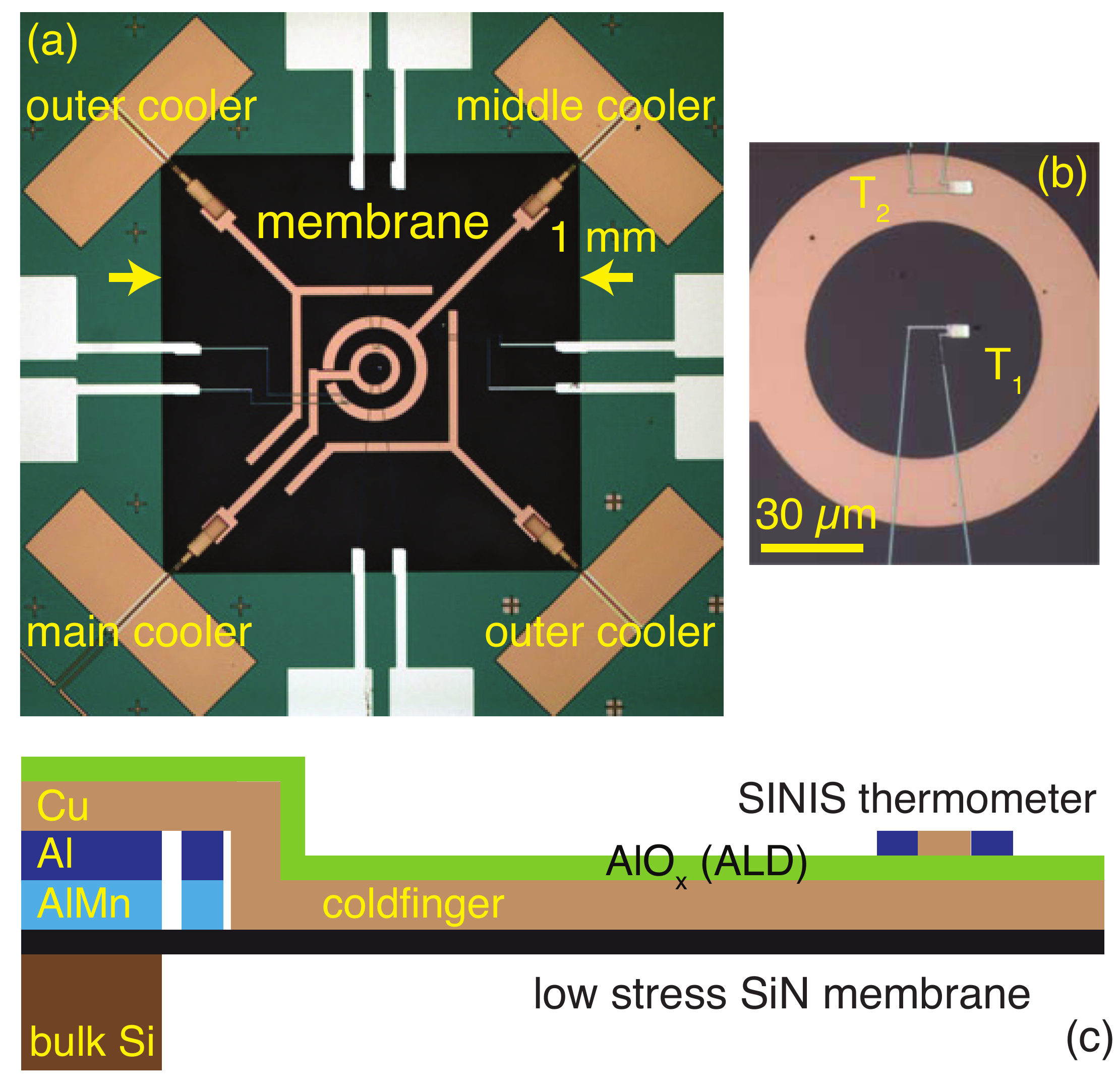}
\caption{(a) Layout of the cooling platform equipped with four SINIS coolers at corners of the low stress SiN membrane. One NIS junction has an overlap area of 200$\times$4 $\mu$m$^2$, and the membrane size is 1$\times$1 mm$^2$. The coldfingers are arranged in such a way that the main coldfinger is surrounded by three others to shunt the leak from the bulk. (b) Zoom of the middle part in (a) that shows a SINIS thermometer at the center of the membrane and another one on top of the coldfinger. (c) Cross-sectional diagram of the device. The cooler with an AlMn quasiparticle drain and a suspended Cu layer is connected to the Cu-only coldfinger that extends to the center of the SiN membrane. The whole device is covered under a thin alumina AlO$_{\text{x}}$ layer. A small SINIS thermometer probes the phonon temperature at the center of the membrane.}
\label{GenFig}
\end{center}
\end{figure}

Following the technique in \cite{NguyenAPL,NguyenNJP}, the fabrication starts by depositing a multilayer on a low stress SiN membrane of size 1 mm $\times$ 1 mm $\times$ 100 nm \cite{Norcada}. Respectively, 200 nm of AlMn, 200 nm of Al, and 60 nm of Cu are sputtered in situ with oxidations in between. The AlMn layer, which is normal \cite{NguyenNJP}, is oxidized in a mixture of Ar:O$_2$ of ratio 10:1 at 2 $\times$10$^{-2}$ mbar pressure for 2 minutes, and the Al layer in pure O$_2$ at 7 mbar for 5 minutes. Two photolithography steps \cite{NguyenAPL} with wet etching of Cu and Al then follow to define the four coolers, which are aligned to corners of the membrane. The size of a NIS junction is 200$\times$4 $\mu$m$^2$ and a cooler has 1 nW cooling power at 300 mK. The normal island of the SINIS has a pair of smaller NIS junctions to probe its temperature, see lower left side of Fig. 1a.

Next, four coldfingers are patterned using electron beam lithography, and 60 nm of Cu is deposited with a lift-off resist scheme. These coldfingers connect the cold normal metal to the center of the membrane. They are designed so that there is one main coldfinger, one middle coldfinger, and two outer coldfingers (see Fig. 3 for different designs). The whole device is then covered by 25 nm of AlO$_{\text{x}}$ using atomic layer deposition (ALD) with H$_2$O and Al(CH$_3$)$_3$ as precursors at pressure 5 mbar and temperature 100$^0$ C. This passive layer isolates the cooler electrically from any structure patterned above it. Besides electron thermometers attached to the cooler itself, SINIS thermometers are placed at various places using electron beam lithography and two angle evaporation. These thermometers are isolated through the alumina layer and read local phonon temperatures. Note that this production step is totally independent from the cooler fabrication and employs state of the art electron beam lithography. Consequently, fabrication of other devices is feasible following this recipe. As we need a normal metal for our SINIS thermometer, the angle of copper deposition is large (50 degree) so that Cu along the connecting lines falls on the wall of the resist and is removed during lift off. The resulting superconducting Al-only wires on the membrane ensure low thermal contact to the outside. During the whole fabrication, temperature is limited to 130$^0$ C to protect the top Cu layer from oxidation as well as to warrant the quality of the tunnel barrier.

The sample is then glued to the stage at the mixing chamber of a dilution cryostat. About 30 gold wires of diameter 25 $\mu$m are bonded around the device from the chip to the sample stage to enhance thermalization of the coolers. Due to the limited number of wires in the cryostat, only the main cooler, which is connected to the centermost coldfinger, is bonded with four wires. The other three coolers are bonded in series and are run independently of the main one. This way, the optimum bias of the whole device is determined by iteratively searching for the optimum of each cooler based on the performance at the center of the membrane of the whole device. This is done automatically using a Matlab script. In other experiments that are not reported here, bonding four coolers either in series or in parallel gave, as expected, similar cooling performance at the center of the membrane. The SINIS thermometers are calibrated to the cryostat temperature when all coolers are unbiased. Throughout this work, bath temperatures of the cryostat are read with a ruthenium oxide resistor calibrated against a Coulomb blockade thermometer \cite{MatthiasIJT11}. The bath temperature is not necessarily equal to the chip temperature (see Fig. 2, diamond symbols), as the latter can be significantly elevated due to the high input power of the cooler of about 100 nW.

\begin{figure}
\begin{center}
\includegraphics[width=3in,keepaspectratio]{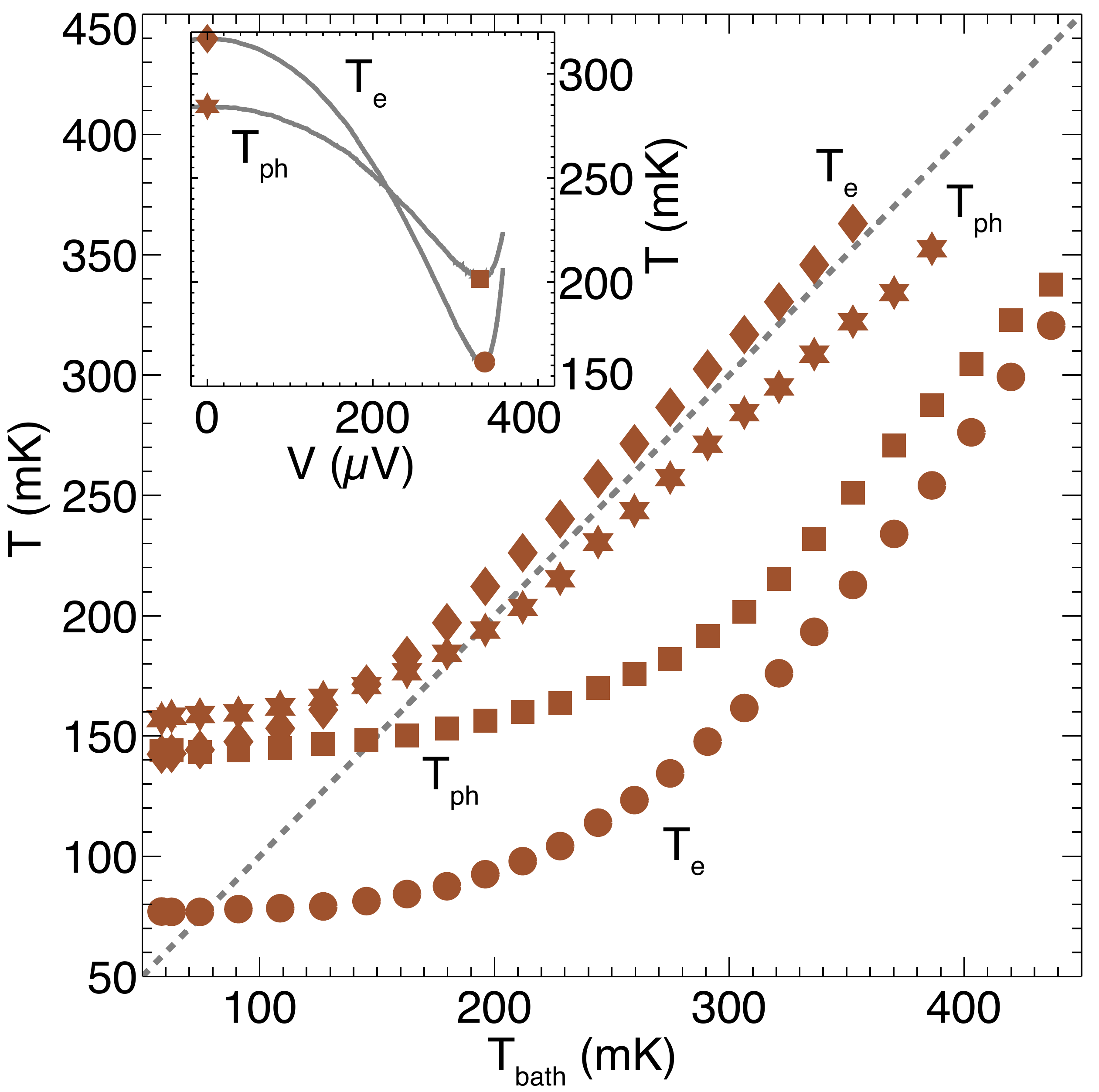}
\caption{Temperatures at different parts of the device as a function of bath temperature when all outer coolers are biased optimally. Diamonds show the temperature of the main cooler at zero bias, which is also the bulk phonon temperature on the chip. Stars show the phonon temperature at the center of the membrane when the main cooler is at zero bias. Squares represent the main result of this work, the lowest phonon temperature on the membrane, and circles are the lowest electron temperature of the main cooler. The dashed line yields $T=T_{\text{bath}}$. The inset shows data at 305 mK bath temperature as a function of bias on the main cooler when other coolers are optimally biased.}
\label{Coola}
\end{center}
\end{figure}

Performance of the device at bath temperature 305 mK is shown in the inset of Fig. 2 when three outer coolers are set to their optimum bias points. $T_e$ shows the electron temperature of the cooler on bulk, and $T_{\text{ph}}$ the phonon temperature at the center of the membrane. At zero bias on the main cooler, $T_e$ reads temperature of the bulk, which is overheated to 317 mK, diamond symbol, due to the high input current $\sim$90 $\mu$A on three other coolers. However, the center of the membrane is cooled to 285 mK, star symbol, due to the three cooled coldfingers, even when the main coldfinger is thermalized to the bath temperature on bulk. At the optimum bias on the main cooler, the center of the membrane is cooled to 200 mK, square smbol. The main figure presents the same data at the optimum working bias versus bath temperature. In general, the chip is always overheated, and the membrane is effectively cooled with extra help from the auxiliary coolers. At optimum cooling, phonon temperature on the membrane reaches 150 mK from 250 mK, which is an achievable base temperature of a $^3$He cryostat. This temperature is quite uniform in this area as the thermometer at the center and the one on top of the colfinger ($T_1$ and $T_2$ in Fig. 1b) show almost the same value (data not shown). Although electrons on the main cooler reach 60 mK similar to the case on bulk \cite{NguyenLowT}, phonons on the membrane saturate at about 150 mK mainly due to the residual heat leak on it and weak electron-phonon interactions in the coldfinger at low temperature ($\dot{Q}_{\text{e-ph}} \propto T_{\text{ph}}^5$).

\begin{figure}[t]
\begin{center}
\includegraphics[width=3in,keepaspectratio]{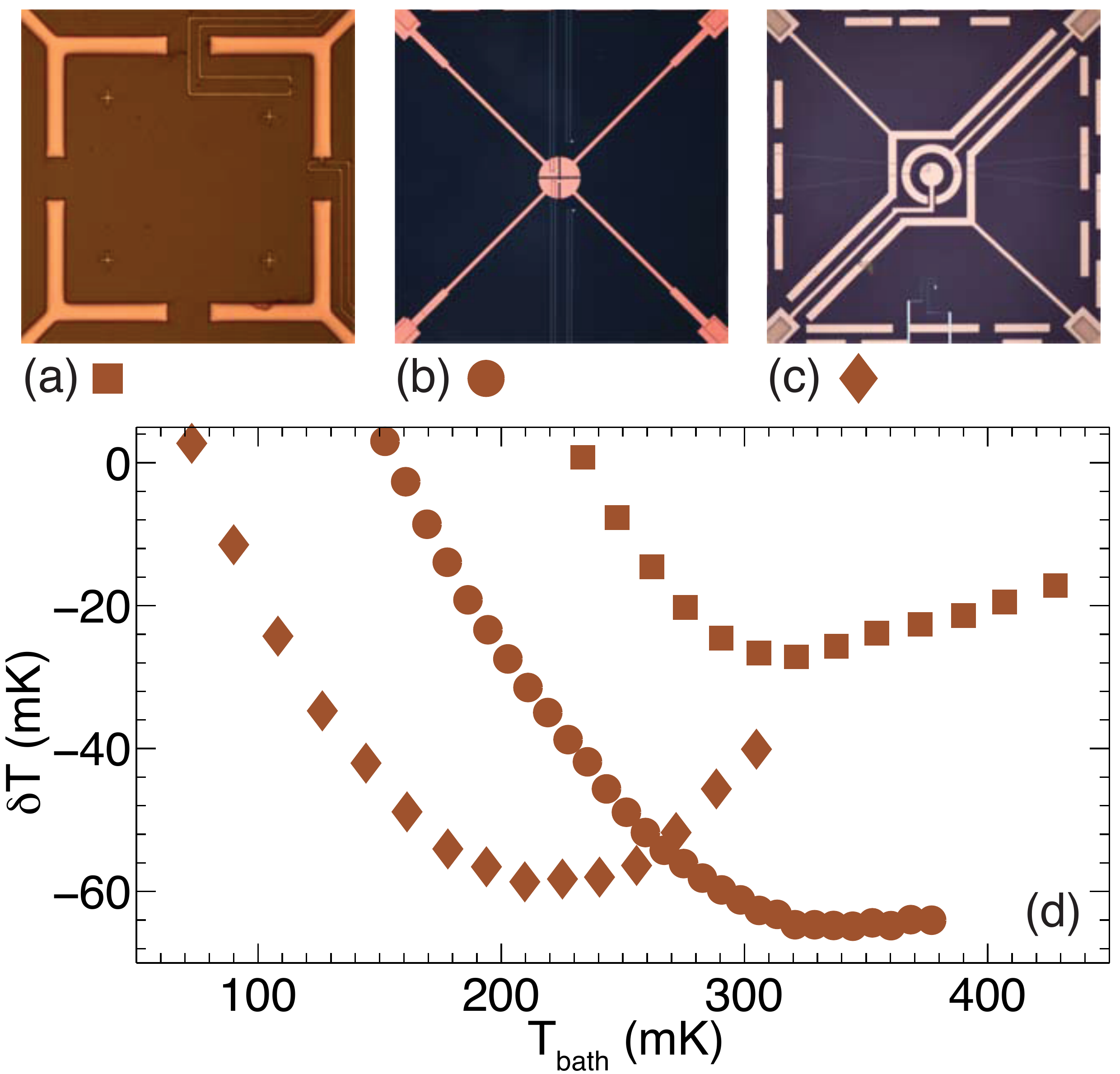}
\caption{(a-c) Different layouts of the coldfinger. The membrane appears as dark color, and the Cu coldfinger as bright color. Each image covers an area of 850$\times$850 $\mu$m$^2$. (d) Phonon temperature at the center of the membrane corresponding to samples (a-c). Note that (c) is similar to that in Fig. 2 but with an inferior performance.}
\label{Coolb}
\end{center}
\end{figure}

We have investigated different layouts of the coldfinger. In all these tests, the membranes are unperforated. A similar design to the device reported in Ref. \cite{MillerAPL08} with a Y shaped coldfinger in Fig. 3a: phonons on the membrane reach 270 mK from 300 mK bath temperature, see square symbols in Fig. 3d. Note that this sample was not covered with AlO$_{\text{x}}$ by ALD and there is a gap between coldfingers for wiring of the thermometers. Apparently, heat leak to the membrane is large. Figure 3b presents another approach, where four coolers play an equal role, and all are connected to a "cold disc" at the center of the membrane. The cold disc should cool the sample on top more efficiently as compared to a piece of membrane that is surrounded by the coldfinger, like the one in Fig. 1a. Separated through the thin alumina layer, the thermometer on top of this cold disc reaches only 240 mK from the bath temperature of 300 mK (circle symbols), eventhough the coolers perform similarly to the one presented in Fig. 2. Figure 3c presents a similar design as the sample of Fig. 1a except that the center part is a cold disc and one cooling junction has an area of 70$\times$4 $\mu$m$^2$. The smaller junction generates less heat on the chip, but provides less cooling power as well. As a result, this particular device works well at low temperatures, but loses its power near 300 mK.

Heuristically, the onion-like coldfinger should work better than the regular design, e.g. the Y-shape used in Ref.\cite{MillerAPL08}. Here, the outer coldfinger precools the inside area and at the same time prevents the external heat load. The inner coldfinger, now precooled to a temperature lower than the bath, can focus on cooling just a small part of the membrane. This is possible because  each cooler has 1 nW power at 300 mK, enough to cool the whole membrane. Of course, it is more desirable to cool the whole inner cooler directly with the outer cooler, similar to a cascade cooler \cite{CourtoisAPL14}. Such a setup requires the superconducting electrodes to be on the membrane, which generates a lot of heat and poses even more challenges.

We have focused on a practical platform that is suitable for a wide range of applications. To start with, it is not perforated. A robust SiN membrane with excellent mechanical properties would survive further integration with other devices. For example, fabrication of a kinetic inductance device or a qubit on it would be quite straightforward following the demonstrated fabrication scheme. Second, the whole platform is passivated with alumina, isolating the foreseen cooled device on top. The passivated alumina layer also allows a free design of the coldfinger layout as required by different applications. These two advances transform the SiN membrane into a robust and versatile platform. Last but not least, the cooler has a high power and can thus accomodate for dissipative devices on top. This platform would provide a phonon bath of 150 mK when attached to a $^3$He adsorption cryostat.

The presented cooler dissipates a large amount of heat in the surroundings of the membrane, up to one microwatt. Although bonding extra gold wires improves thermalization, as reflected on the improved phonon temperature of the membrane (data not shown), this is kind of a hasty solution to the on-chip overheating issue. A more systematic approach of the whole set up is desired, such as using gold ribbon, and clamping the device to the sample stage. Moreover, each cooler should focus on its own task. The outer cooler should aim at high power to shunt the heat leak, and the inner cooler should seek for high performance, either by reducing the size of the cooling junction or by increasing the tunnel barrier thickness for a weaker overheating of the superconducting electrode \cite{NguyenLowT}. Additionally, other material choices, such as nano-perforated \cite{HeathNnano, MaasiltaNComm14}, nano-laminated \cite{TalghaderJAP11,GeorgeSci04}, or corrugated membranes \cite{Sainiemi,Berdova} might also enhance thermal isolation between the cold and the hot regions. Finally, the cryostat used for the present measurement had a small power that was barely enough to support the high input current of the device (200 $\mu$A). 

In conclusion, we have transformed a standard commercial SiN membrane into a cooling platform by integrating it with powerful SINIS refrigerators. Phonons at the center of the membrane reach 150 mK from 250 mK starting temperature, the base temperature of a $^3$He cryostat. This passivated, unperforated platform is compatible with a wide range of requirements set by practical implementations.

We thank H. Courtois for discussions, A. Peltonen for studies of the membrane, and M. Berdova and S. Franssila for supplying other membranes besides SiN for tests. We acknowledge the support of the European Community Research Infrastructures under the FP7 Capacities Specific Programme, MICROKELVIN project number 228464, the EPSRC grant EP/F040784/1, the Academy of Finland through its LTQ CoE grant (project no. 250280), and the Otaniemi Research Infrastructure for Micro and Nano Technologies. Samples were fabricated in the Micronova Nanofabrication Center of Aalto University.

\end{document}